# Trust in Data Science: Collaboration, Translation, and Accountability in Corporate Data Science Projects

SAMIR PASSI, Cornell University, USA
STEVEN J. JACKSON, Cornell University, USA

The trustworthiness of data science systems in applied and real-world settings emerges from the resolution of specific tensions through situated, pragmatic, and ongoing forms of work. Drawing on research in CSCW, critical data studies, and history and sociology of science, and six months of immersive ethnographic fieldwork with a corporate data science team, we describe four common tensions in applied data science work: (un)equivocal numbers, (counter)intuitive knowledge, (in)credible data, and (in)scrutable models. We show how organizational actors establish and re-negotiate trust under messy and uncertain analytic conditions through practices of skepticism, assessment, and credibility. Highlighting the collaborative and heterogeneous nature of real-world data science, we show how the management of trust in applied corporate data science settings depends not only on pre-processing and quantification, but also on negotiation and translation. We conclude by discussing the implications of our findings for data science research and practice, both within and beyond CSCW.

CCS Concepts: • **Human-centered computing** → **Collaborative and Social Computing** → **Empirical Studies in Collaborative and Social Computing**

**KEYWORDS**
Trust; Credibility; Data Science; Collaboration; Organizational Work

## 1 INTRODUCTION

Data science tools and techniques constitute an increasingly common and complex feature of contemporary work settings, impacting modes of knowledge and governance [27,66,67], shaping organizational practices [15,52,99], reconfiguring publics [20,50,56], and simultaneously enabling and constraining the possibilities of human action and agency [9,10,74].[1] But the limits and tensions of data science work in its collaborative dimensions are not yet fully scoped. Researchers argue that "everything might be collected and connected, [but] that does not necessarily mean that everything can be known" [80]. Data are never "raw" [34], often speaking specific forms of knowledge to power [10,74,102]. In the current push towards algorithmic analyses, what counts as

---

Authors' addresses: {sp966, sjj54}@cornell.edu, Department of Information Science, Cornell University.

[1] The term 'data science,' as used in this paper, refers to the collective practices of assembling, organizing, processing, modeling, and analyzing data towards specific goals through computational and statistical techniques drawn from domains such as machine learning, data mining, natural language processing, and artificial intelligence.

Pre-print version.

valid and reliable knowledge remains contested [52–54,63]. Algorithmic results based on "sufficient" data, however, are often considered "credible" enough to serve as actionable evidence [43,104]. "[The] idea of data-led objective discoveries entirely discounts the role of interpretive frameworks in making sense of data which [are] a necessary and inevitable part of interacting with the world, people and phenomena" [69:320]. If data science enables us to ask different questions and generate novel insights, it also requires varied and distributed forms of interpretation and discretionary judgment that ensure meaningfulness, confidence, and reliability of algorithmic results. Put simply: data science is a sociomaterial practice [73] in which human and technical forms of work intertwine in specific, significant, and mutually shaping ways. The effective practice and governance of data science therefore remain top priorities for data science researchers and practitioners alike.

Central to all of this is trust. The credibility of algorithmic results is adjudged variously—at times through statistical probabilities and performance metrics, and at other times through prior knowledge and expert judgments. The reliability and valence of algorithmic results, as data scientists well know, is also impacted by factors such as the framing of questions, the choice of algorithms, and the calibration of models.[2] Established credibility mechanisms such as statistical significances, quantified metrics, and theoretical bounds comprise forms of *calculated trust*, but their effective use remains challenging. Data scientists may have the necessary expertise to test systems, but the manual intractability of large-scale data coupled with the complex, sometimes opaque, nature of state-of-the-art models makes it hard even for them to clearly articulate and ascribe trustworthiness to approaches and insights. This is even harder for lay users who lack specialized data science knowledge but are sometimes the users of these systems or those most impacted by them. CSCW and HCI researchers have begun to take on these challenges through efforts to make results more understandable [81,82], to effectively manage data science systems [55,65], to understand user perception of performance metrics [48], and to ascertain ways to foster trust through system design [57,58].

Contemporary understanding of data science in research is largely based on the analysis of data science work in academic and research sites [26,66,71,75,76,83,98]; shaped by limits of access, confidentiality, and non-disclosure, the large body of applied data science work in corporate settings has received much less attention. Crawford & Calo [19] argue that the lack of research focus on already in-use data science systems has created a "blind spot" in our understanding of data science work. Based on ethnographic fieldwork in a corporate data science team, this paper attempts to bridge this gap by describing how problems of trust and credibility are negotiated and managed by actors in applied and corporate settings, with a focus on two separate projects: *churn prediction* and *special financing*. We describe how four common tensions in corporate data science work – (un)equivocal numbers, (counter)intuitive knowledge, (in)credible data, and (in)scrutable models – raise problems of trust, and show the practices of skepticism, assessment, and credibility by which organizational actors establish and re-negotiate trust under uncertain analytic conditions: work that is simultaneously calculative and collaborative. Highlighting the heterogeneous nature of real-world data science, we show how management and accountability of trust in applied data science work depends not only on pre-processing and quantification, but also on negotiation and translation—producing forms of what we identify as 'algorithmic witnessing' and 'deliberative accountability.' Trust in data science is therefore best understood as a deeply *collaborative* accomplishment, undertaken in the service of pragmatic ways of acting in an uncertain world.

---

[2] The term 'algorithm' refers to the underlying statistical/computational, approach (e.g., 'random forests' is an algorithm). The term 'model' refers to the specific application of an algorithm to a dataset (e.g., using 'random forests' on 'user data' produces a model).

In the following sections, we begin by reviewing history and sociology of science, social science, critical data science, and CSCW literature on trust and credibility in science and technology more broadly and data science and organizations more specifically. We then describe our research site and methodology, before moving to two empirical examples illustrating the challenges and complexity of trust in applied data science work. We conclude by describing our findings concerning the negotiation of trust in real-world data science work, highlighting the implications of our findings for the data science field, both within and beyond CSCW.

## 2 TRUST, OBJECTIVITY, AND JUSTIFICATION

Our conceptualization of trust begins with an important vein of work in history and sociology of science on the relation between science, trust, and credibility. Influential work by Shapin [88] and Shapin & Schaffer [91] show how "working solutions" to problems of credibility in early experimental science were found in the social perception of gentlemen as "reliable truth-tellers." A gentleman was a person of noble birth, raised in the traditions of civility and therefore marked by a commitment to honor, virtue, and righteousness. The seventeenth-century cultural association of "gentility, integrity, and credibility" provided forms of *social scaffolding* to negotiations of scientific trust—gentlemen rejected "notions of truth, certainty, rigor, and precision which were judged suitable for scholarly inquiry, but out of place in civil conversations" [88:xxx]. The perception of gentlemen as oracles of truth, however, rendered other knowers such as laboratory technicians invisible [87]. Early experimental science was embedded in (and built on) a "moral economy of truth" led by gentlemanly scientists pursuing science with "epistemological decorum." At the same time, techniques of "*virtual witnessing*" – organized around mechanized experiments, standardized (in principle) reproducible methods, and conventionalized forms of reporting – helped discipline the senses and lent certainty to experimental knowledge. This combination of social order and mechanical apparatus working together – and not uncredited and single testimonies – instilled trust and ultimately power in experimental results. Early experimental science was thus a collective practice simultaneously social and technical: facts emerged through specific forms of sociability embedded within the experimental discourse.

It was only towards the mid-nineteenth century that "objectivity," as we understand it today, gained prominence as a central scientific ideal [22,23]. Daston & Galison [23], for instance, describe a shift from "truth-to-nature," a mode of objectivity characterized by "reasoned images" of natural reality produced by scientists based on theoretical selection and aesthetic judgments of exemplary specimens to "mechanical objectivity," organized around the credibility of mechanized means such as photography to produce images untouched by (troubling) individual scientific judgments. As the work of interpretation shifted from the maker to the reader, scientific artifacts became open to interpretation, making consensus challenging. Agreements between multiple ways of seeing required differentiating between right and wrong interpretations: science required the correct "professional vision" [36]. As mathematicians and physicists argued for "structural objectivity" characterized by forms of measurement and replicability, the role of "trained judgments" became salient. Scientists thus chased truth with "calibrated eyes" and standardized tools, simultaneously questioning the credibility of their findings, instruments, and knowledge.

Today, objectivity goes hand in hand with quantification: mechanically-produced numbers standing in for factual representations of reality [24,28,47,72,77]. Numbers do not just lend credibility but are themselves "couched in a rhetoric of factuality, imbued with an ethos of neutrality, and presented with an aura of certainty" [83:4]. Numbers ascribe but also command trust. Trust in numbers, however, as Porter [77] argues, is but one form of "technologies of trust." Through studies of accounting and engineering practices, Porter shows how the "authority of numbers" was enacted as a pragmatic response to institutional problems of trust. The twentieth-century thrust for precision coupled with the rise of mechanical objectivity brought in a "regime of calculation" in which rule-bound numbers fostered a "cult of impersonality." Sanitizing the inherently discretionary nature of professional practices, quantification became a rigorous judgment criterion: "a way of making decisions without seeming to decide" [77:8]. Numbers, however, don't contain within themselves their own logic of interpretation. As a form of intervening in the world, quantification necessitates its own ecologies of usability and valuation. Trust in numbers is therefore best understood as a variegated "practical accomplishment" [31,32], emanating from efforts at standardization and mechanization, along with forms of professional and institutional work [40,42,78,89,90].

A second line of work central to problems of trust in complex organizational settings is found in pragmatist traditions of social and organizational science. Dewey [25] argues that instead of existing as *a priori* criteria, values – as perceived or assigned worth of things – are continuously negotiated within decision-making. Processes of valuation are simultaneously evaluative (how to value?) and declarative (what is valuable?). Boltanski & Thévenot [8] focus on the "plurality of justification logics" comprising valuation practices. They describe six orders of worth – *civic, market, industrial, domestic, fame,* and *inspiration* – within which "justifiable agreements" are achieved in social and professional practices. Each order of worth signifies its own reality test based on "the testimony of a worthy person, […] the greatest number, […] the general will, […] the established price, […or] the competence of experts" [8:132]

Applying these insights to organizational decision-making, Stark [94] builds on Boltanski & Thévenot's [8] work to build a theory of "heterarchies"—organizational forms in which professionals operate according to diverse and non-commensurate valuation principles. In his ethnographic studies of a factory workshop, a new media firm, and an arbitrage trading room, Stark analyzes how actors use diverse "evaluative and calculative practices" to accomplish practical actions in the face of uncertainty. In heterarchies, decision-making requires negotiations between different, often competing, performance metrics. Stark distinguishes his account of heterarchies from Boltanski & Thévenot's orders of worth in two significant ways. First, while Stark argues that decisions are embedded within multiple forms of valuation, he does not believe that decisions are confined within a pre-defined matrix of worth. Valuation criteria, for Stark, remain contextual, emerging differently in different situations. Second, while Boltanski & Thévenot position orders of worth as ways of *resolving* uncertainty, Stark sees the plurality of valuation as providing opportunities for action by *creating* uncertainty. Organizational conflicts are not roadblocks, but integral to organizational diversity in which the "productive friction" between multiple ecologies of valuation helps accomplish justification and trust—(dis)trusting specific things, actions, and worlds.

Data science is no different. Integral to several contemporary knowledge practices, algorithmic knowledge production is here and now. Big data pushes "the tenets of mechanical objectivity into ever more areas of applications"—data science is not just interested in calculating what *is,* but also "aspires to calculate what is yet to come" [83:4]. Given enough data, numbers appear to speak for themselves, supplying necessary explanations for all observations [3,43,62,83]. The problem is exacerbated given that automation – as operationalized in and through data science – is often understood as an absence of bias [12,27,33,70]. The mathematical foundation of algorithms, coupled with the manual intractability of large-scale data, has shaped the use of quantified metrics as key indicators to ascertain a data science system's workability.

The increased management of data science systems is a priority [65,96,105], but "the complex decision-making structure" needed to manage them often exceeds "the human and organizational resources available for oversight" [68]. One limiting factor is the opacity of data science systems [97,100] arising from a variety of reasons: (a) algorithms are often trade-secrets, (b) data science requires specialized knowledge, and (c) novel analytic methods remain conceptually challenging [14]. Neural networks are often critiqued for their black-boxed nature,[3] but researchers argue that even simpler models are not necessarily more interpretable than their complex counterparts [64]. The fact that models do not always furnish explanations has sparked efforts to produce humanly understandable explanations of data science workings and results [38,82,92]. Another limiting factor is the paywalled or proprietary nature of datasets. Several data correlations established in large datasets are neither reproducible nor falsifiable [45,61]. The combination of data inaccessibility and analytic fragility makes "virtual witnessing" [91] challenging, if not at times impossible. While making data public may seem a good first step (though not in the best interests of corporations), transparency remains a difficult and problematic ideal [2]. Data science comprises myriad forms of discretionary human work—visible only to those in its close vicinity. Algorithmic results comprise specific articulations of "data vision"—the situated rule-*based* traversal of messy data through clinical rule-*bound* vehicles of analysis [75]. The absence of development histories of systems further camouflages aspects arising not only from personal and professional values, but also from the emergent sites of algorithmic re-use as opposed to their contexts of development [29,43,51]. This combination of oversight challenges often causes "non-expert" data subjects to lose trust in the enterprise of data science more broadly [17,27,84].

CSCW and critical data science researchers have thus put forward an agenda for "human-centered data science," arguing that focusing on computational or statistical approaches alone often fails to capture "social nuances, affective relationships, or ethical, value-driven concerns" in data science practices [4:2]. Such failures cause rifts between developers' understanding of what the system does and the users' interpretation of the system in practice, causing user and societal confidence in these systems to plummet. This is particularly challenging because the systems are not usually designed "with the information needs of those individuals facing significant personal consequences of model outputs" [7:10]. While quantification provides calculative relief in the face of uncertainty, questions of trust are "ultimately questions of interpretation," and within meaning-making processes "limitations of traditional algorithm design and evaluation methods" are clearly visible [6:10]. Researchers thus focus not only on fostering trust in computational systems [58], but also on understanding users' perception of quantified metrics [48].

These bodies of history and sociology of science, critical data studies, CSCW, and social science research highlight the import of trust and credibility justifications in data science work. Problems of trust in science are solved variously, characterized by a set of key approaches: *social scaffolding*, *virtual witnessing*, *mechanical objectivity*, *trained judgments*, and *quantification*. Beyond a taxonomy of trust mechanisms, this list also makes visible the sociocultural and political factors encompassing scientific negotiations on trust. As we grapple with problems of societal trust in science and technology, it is important to understand how trust and credibility are established in scientific and technological practices at large. Data science charts new knowledge frontiers, but its maps remain opaque and distant to all but a few. In complex organization settings, data science is transected by multiple experts, interests, and goals, relying upon and feeding into a plethora of practices such as business analytics, product design, and project management. Applied data science needs not only scientists and engineers, but also managers and executives [55,65].

---

[3] A neural network contains multiple layers with each layer containing several nodes. Broadly, each node acts as a simple classifier, activating upon encountering specific data attributes. Output from one layer forms the input for the next layer.

In this paper, we unpack the situated work of organizational actors to engender trust in data, algorithms, models, and results in corporate data science practices. Highlighting the plurality of valuation practices in organizational decision-making, we describe common tensions in data science work and mechanisms by which actors resolve problems of trust and credibility. As forms of justification, these mechanisms exemplify pragmatic, not absolute, forms of trust, helping actors act and move forward in an uncertain world—valuing specific forms of knowing over others. Our goal in this paper is not to merely argue the partial, social, and messy nature of data, models, and numbers (a fact readily acknowledged by our actors), but to show *how* actors negotiate and justify the worth of data science solutions to identify opportunities for practical action when faced with such conditions. Evaluation of a data science system, as we show in this paper, is rarely the simple application of a specific set of criteria to a technology, and often requires specific mechanisms to negotiate and translate between multiple situated, often divergent, performance criteria. In this paper, we address two of these mechanisms: *algorithmic witnessing*, in which data scientists assess model performance by variously, mostly technically, reproducing models and results; and, *deliberative accountability*, in which multiple experts assess systems through collaborative negotiations between diverse forms of trained judgments and performance criteria.

## 3 RESEARCH SITE, METHODS, AND FINDINGS

The paper that follows builds on six months of ethnographic fieldwork with DeepNetwork[4], a multi-billion-dollar US-based e-commerce and new media organization. To gain more immersive and participatory access to ongoing and ordinary work practices and experiences, the first author worked as a data scientist at the organization between June and November 2017, serving as lead data scientist on two business projects (not reported in this paper) and participating in several others. Founded in the nineties, DeepNetwork owns 100+ companies in domains such as health, legal, and automotive. A number of these are multi-million-dollar companies with several thousand clients each. The organization has a core data science team based on the US West Coast that works with multiple businesses across different domains and states. There are multiple teams of data engineers, software developers, and business analysts both at DeepNetwork and its subsidiaries. While not a research-driven data science organization (in the fashion, for example, of Google, Amazon, or Facebook), DeepNetwork remains an extremely data-driven company with copious amounts of data across numerous businesses. Its focus however is not data science *research* (though it holds data science patents), but *business applications* of data science. During our time at the company, the data science team comprised eight to eleven members (including the first author). The team's supervisor is Martin—DeepNetwork's Director of Data Science with 30+ years of industry experience managing business projects in several major technology firms. Martin and the data science team report directly to Justin—DeepNetwork's Chief Technology Officer (CTO).

---

[4] All organization and personnel names have been replaced with pseudonyms.

Justin is one of the company's seven executives and has 20+ years of experience in the technology industry.

The first author applied for the data scientist position at DeepNetwork, going through a series of technical and behavioral interviews. All interviewees were explicitly informed that our primary goal was to research corporate data science practices. We clearly stated that we would work at the organization for a fixed time duration (three to six months) and would need explicit permission to conduct research in company premises, gathering data such as field notes and audio-recorded interviews. DeepNetwork's decision to hire the first author as an interning data scientist was based primarily on the assessment of the first author's data science knowledge and expertise. As part of the negotiation process, we settled on three key aspects of our research design. *First,* participation was optional, and each participant needed to sign a consent form (we provided a copy of our consent form for approval). Participants could opt-out of research at any point, and all company, personnel, and project names would be replaced with pseudonyms to preserve anonymity and privacy. *Second,* participants could consent to selective participation. E.g., a participant could consent to the fieldwork (i.e., have their data recorded in field notes) but not to the interview. Participants could further choose not to have their interviews audio-recorded (we provided copies of our interview topic guides for approval but explained that the guides would change over the course of research). *Third,* the organization could define what research data could and could not be collected, but the analysis of ethnographic and interview data, including for reasons of inter-participant privacy, would be done solely by the researchers. We also provided a copy of the approval certificate for our research project from our university's Institutional Review Board (IRB). All our submitted documents were vetted by DeepNetwork's Human Resources and Legal departments. DeepNetwork agreed to research participation. As part of the non-disclosure agreement, we agreed not to make any copies of company datasets and proprietary code. In practice, both participants and company leaders participated willingly and openly in research, and only two people declined our request to record their interviews.

During the six-month period, we conducted 50+ interviews with data scientists, project managers, product managers, business analysts, and company executives and produced 400+ pages of fieldwork notes and 100+ photographs. Interviews and fieldnotes were transcribed and coded according to the principles of grounded theory analysis, as articulated by Anselm Strauss and students [16,95] and as previously applied in CSCW research by scholars such as Ellen Balka [5,85] and Susan Leigh Star [93]. During fieldwork, we began to encounter discrepancies between how different organizational actors (such as data scientists, project managers, and business analysts) articulated problems with and confidence in data and models. We saw how data science projects consisted of diverse negotiations to tackle varying forms of uncertainties. We followed up on this theme during fieldwork by focusing our attention on the points of friction and collaboration between different organizational actors and groups (in field notes and interviews). Post-fieldwork, we organized our interview and fieldwork data in two ways: (a) categorized by projects (e.g., all data on the project *churn prediction* in one place), and (b) categorized by actor groups (e.g., all data on *business analysts* in one place). The former enabled us to analyze themes within and across different projects, and the latter allowed us to examine the perspectives of specific professional groups.

The theme of trust began to emerge, as part of our own thematic coding, when analyzing the data about forms of friction and collaboration between different organizational actors. We identified a series of situated tensions in corporate data science work (the four most salient of which we report below) through two rounds of coding (with open and in-vivo codes[5] such as explanation, translation, expertise, intuition, mess, and verification). Per grounded theory principle of 'constant comparison' [35], we frequently juxtaposed and analyzed individual tensions across different corporate project and professional group descriptions looking for both distinctions and common patterns. We started to see in the situated resolutions of these tensions a common recourse to making (in)visible specific aspects of data science work. For example, the incompleteness of datasets was often foregrounded as a key reason for sub-optimal model performance in project meetings. Our subsequent analysis of the strategies that actors used to resolve these tensions comprised a second round of coding with open and in-vivo codes such as testing, data processing, value, and narrativization. We chose trust as the organizing theme for this paper, but in our analysis the theme of trust remained (and remains) intermingled with several other themes not explicitly addressed in this paper, including standardization, invisible work, materiality, and repair. While the specific project cases reported in this paper were selected for their salience to the tensions and practices highlighted here, they are broadly representative of patterns identified across the much larger number of cases not reported on and are moreover consistent with the first author's experience as project lead or collaborator on separate projects not reported in this paper. Specific quotations in the empirical stories that follow reflect a mix of transcript data from project meetings (and thus embedded in the ordinary flow of everyday corporate work), and separate interviews conducted with the first author. Audio-recorded personal communications are referenced as 'interviews' and non-audio-recorded conversations from team meetings and everyday work are referenced as 'fieldwork notes.' We have provided short biographical notes on each actor to provide further empirical context (certain details have been omitted for participant anonymity).

The first author's dual role as both ethnographer and data scientist, communicated explicitly in the first meeting with each participant, presented both research opportunities and challenges. On the one hand, working as a data scientist helped build rapport with corporate personnel. For instance, after about four weeks, everyday meetings between the first author and data science team members were dominated by detailed and practical conversations around technical challenges, algorithmic issues, and ongoing projects. Even non-data-scientists began to see the first author primarily *as* a data scientist—in audio-recorded interviews that were mainly organized for research purposes, for instance, product managers and business analysts would continue to ask questions about and discuss project updates and requirements. The fact that after about two months the first author became the lead data scientist on two projects further helped solidify their primary identity as that of a data scientist—to the extent, for instance, that the data scientists often explicitly articulated surprise (e.g.: 'but, you knew that, right?') when the first author would ask them to explain aspects of their everyday work for the ethnographic record. On the other hand, the first author's data scientist identity posed difficulties in personal communication with non-data-scientists. Business team members, for instance, were hesitant to point out mistakes and problems with the data science team, sometimes resorting to conciliatory descriptions such as 'your team did its best,' 'maybe the problem is on our side,' and 'we do love the work your team is doing.' In these moments, the first author had to work to make visible their dual role, reminding the interviewees that this was after all a research interview and assuring them that their critique would never be directly communicated to CTO Justin, director of data science Martin, or other data scientists.

---

[5] In-vivo coding involves generating codes *from* the empirical data (e.g.: using an actor phrase as a code) while open coding involves codes manually assigned by the researcher based on ongoing analysis of the empirical data.

### 3.1 CASE 1 | CHURN PREDICTION

For case one, we focus on a *churn prediction* project – i.e. the detection of currently active customers who are likely to cancel paid services in the future – associated with DeltaTribe, a multi-million-dollar marketing and management subsidiary owned by DeepNetwork, that provides online customer management services to thousands of clients across the United States in domains such as medical, dental, veterinary, and automotive. In business terminology, customers who cancel paid services are called 'churned,' and active customers who might cancel paid services are called 'likely to churn.' In what follows, we describe key moments from the project to show how tensions emerge as actors negotiate the trustworthiness of numbers and intuition.

**(Un)equivocal Numbers**

When we began fieldwork, the project's data scientists – David[6] and Max[7] – had already settled on two models after trying several approaches.[8] Their models differed not only in their algorithmic approaches, but also in their data. David's model used *subscription data* consisting of information such as the services enabled for a customer and their platform settings. Collected monthly, this data was available for the past two years—a total of 24 datasets. Max's model used *action data* containing data on customers' actions such as the modules they accessed and when. This data was collected every day but was not archived—each day's collection overwrote the previous day's data. David's subscription-based model therefore had more historical data to work with than Max's action-based model. David's model, however, had lower accuracy than Max's model.[9] Both believed that "more data was better," but the jury was out on whether more data led to better performance. In fact, Max's model had an accuracy of ~30%—a number higher only in comparison to the ~20% accuracy of David's model. Director of data science Martin[10] was "disappointed" in both models.

Over time, accuracy scores did not increase, but the data science team's disappointment in their models decreased. This was because accuracy score – as a performance metric – was devalued by team members over the course of the project. For instance, data science team's project manager Daniel[11] found accuracy scores problematic given his prior experiences of working with business teams:

> **Daniel (project manager):** When we say to business teams that we have 90% accuracy, they think: 'oh, you are saying with a 90% confidence that these people are going to churn.' NO, THAT'S NOT WHAT WE ARE SAYING! They think in black and white: likely to churn or not likely to churn. In total we have four answers: (a) people we say will churn, and who churn [true positives], (b) people we say will churn, but don't [false positives], (c) people we say won't churn, and don't [true negatives], and (d) people we say won't churn, but churn [false negatives]. Business folks don't know what the numbers *really* mean (Fieldwork Notes, June 8, 2017).

---

[6] David has an undergraduate degree in electrical and electronics engineering and several online data science certifications. He has been with DeepNetwork for a little less than two years. Before this, he worked as a Product Engineer.
[7] Max has an undergraduate degree in information and computer science and a graduate degree in mathematics and statistics. He has been with DeepNetwork for a little less than two years. Before this, he worked as a Statistical Consultant.
[8] Model one was based on gradient boosting algorithm (xgboost), and model two on random forests.
[9] In churn prediction, the accuracy score refers to the percentage of correctly identified likely-to-churn customers (who do cancel their services in the future) in the total number of likely-to-churn customers identified by a model.
[10] Martin has an undergraduate degree in electronics and electrical engineering, and graduate degrees in computer science and business administration. He has been with DeepNetwork for 5+ years. Before this, he had 25+ years of experience in the technology industry in several roles such as vice-president engineering, director engineering, and research engineer.
[11] Daniel has an undergraduate and a graduate degree in computer science. He worked at DeepNetwork for 2.5 years (he left in late 2017 to join as a process manager in an academic institute). Before this, he worked as a Technology Consultant. Outside the technological domain, he has eight years of work experience in industrial warehousing and public relations.

This point was borne out in a subsequent interview with Parth[12], DeltaTribe's business analyst, who described his perception of low accuracy scores:

> **Parth (business analyst):** "Expectation-wise […], this project was going to be the silver bullet. Then, expectations were, I guess, neutralized or you know, brought down to earth. […] I thought [it] was going to be 80% or 90% accuracy. […] In reality, it was more like 20-30%. That's where the expectations for me were shattered, I guess" (Interview, October 26, 2017).

For Parth, a low accuracy score signaled failure. His assessment was largely based on performance numbers that he believed provided definite information about the model. Both Parth and Charles[13] – another business analyst – were also skeptical of numbers produced by the models: the likely-to-churn customer probabilities. A likely-to-churn probability, generated by the model, indicates the model's perception of a customer's likelihood to churn (the higher the number, the higher the likelihood). Parth and Charles found the probabilities helpful but incomplete on their own without further information on how they were generated.

> **Charles (business analyst):** "We were looking for [likely-to-churn probabilities]. But, keep in mind that, that number does not tell the full story. It is a good indicator, but it is not the absolute truth for us. […] It [only] helps us identify which customers to go for, [but not why]."
>
> **Parth (business analyst):** "The more we understand how models work […] allow[s] us to understand: 'ok, this score actually means this or when there is a change, [it] means that'." (ibid.)

These probabilities were not always a part of the results. Initially, the results only contained customer IDs and a label indicating whether the model perceived a customer as 'likely-to-churn' or 'not-likely-to-churn.' The move to probabilities, as we show below, was largely an outcome of the devaluation of accuracy scores. In addition to voicing concerns about the business interpretation of accuracy scores, and performance metrics in general, project manager Daniel and data scientists David and Max argued that low scores in the project were not a sign of "bad models," but of the "extremely hard" nature of churn prediction itself.

> **David (data scientist):** It is tough. Customers churn for any reason and you can't know that. E.g., they want to save money, or their boss told them to cancel. Even with all the data, we can't be perfect. Almost impossible (Fieldwork Notes, August 16, 2017).
>
> **Daniel (project manager):** "[Accuracy] is so low because you are trying to predict a *very* difficult concept, […] to explain human behavior. […] It is random to a degree. There is stuff that you can predict and model, but […] it just seems unreasonable to […] a data scientist that you can create a model that perfectly models human behavior. […] The challenge with non-technical people is they think that computers can do more than they really can" (Interview, June 30, 2017).

Most predictions made by the models were incorrect, but the data science team gradually saw value in the small handful of correct predictions: *30% was better than nothing.* As Daniel put it, "even bad results or less than ideal results can be good—it is more than what you know now." The perceived value of having *some* predictions over *none* further shaped the data science team's outlook towards model results and metrics. This prompted the move towards likely-to-churn probabilities. While binary labels – likely-to-churn/not-likely-to-churn – drew attention to which labels were right and which were wrong, churn probabilities instead brought to light customers at varying levels of churn risk. The results were now in the form of an ordered list of customers with decreasing likely-to-churn probabilities.

---

[12] Parth has been with DeltaTribe for 8+ years. Before this, he was an Assistant Manager in a corporation (4+ years).
[13] Charles has been with DeltaTribe for 5+ years. Before this, he was a Consultant in a corporation (5+ years).

> **Martin (director of data science):** Think of it as a list that if you are on it, you are going to die. We provide the ones with the highest probability of dying and tell our business team that maybe you should reach out to these folks. By doing this we deemphasize the 30% and say: 'how can a low 30% score still be made useful for businesses?' (Fieldwork Notes, May 30, 2017).

**(Counter)intuitive Knowledge**

The results were presented to DeltaTribe's business analysts Parth and Charles. A feature importance list accompanied the results.[14] Director of data science Martin and project manager Daniel considered the list an important part of the results—an explanation mechanism to help the business team "see" what the models considered noteworthy. Within minutes of seeing the results, both Parth and Charles described the results as "useful." Their perception of the usefulness of results, however, was not grounded in model performance metrics (mentioned once in the hour-long meeting) or in the manual inspection of customer accounts (not done in the meeting at all). Instead, their perception was based solely on the feature importance list.

> **Daniel (project manager):** Here [points to a slide with feature importances] we see that whether customer enabled email and voice were important. So was use, how much the customer engages. It might mean engagement with the platform is good, and customers who engage remain active. Also [points to a different part] communication was important.
>
> **Martin (director of data science):** Do these results match your gut feeling, your intuition?
>
> **Charles (business analyst):** Definitely! These are things we already focus on (Fieldwork Notes, June 9, 2017).

Certain highly-weighted features matched business intuitions, and everyone in the meeting considered this a good thing. Models that knew "nothing about the business" had correctly identified certain aspects integral to business practices. Such forms of intuitive results were important not only for business analysts, but also for data scientists. In an interview, director of data science Martin described how "seeing" intuitive patterns had helped him corroborate results:

> **Martin (director of data science):** "We saw patterns [showing that] there is this cohort of customers that are less than one year old, and they are clustered in their behavior. [...] This other cluster of customers that have been with us more than X number of years and [...] their behavior looks like this. [...] The graph was SO important. [...] To me, it is easier to look at a graph, especially one [...] so *obvious* in terms of patterns" (Interview, August 23, 2017).

In the meeting with business analysts, however, not all feature importances mirrored business expectations. Parth asked Daniel why a specific feature that DeltaTribe considered important did not show up in the top ten features identified by the models. Daniel argued that even if expected features were absent, it did not mean that the models were necessarily wrong:

> **Daniel (project manager):** If you see a feature in the middle of the pack but expect it to be higher, it might mean that in your business you already focus on it. Its importance perhaps went down over time. If you focus on these [points to top features] that are prioritized by models, we expect that these will also go down over time. You focus on customers that we say are at risk. They don't cancel. This is good. But, it means that features will change (Fieldwork Notes, June 9, 2017).

Regarding counter-intuitive feature importances, Daniel reminded Parth and Charles that machine-learning models do not approach data in the same way humans do. He pointed out that models use "a lot of complex math" to tell us things that we may not know or fully understand. While DeltaTribe may consider a feature important for their business, models may spot the over-time devaluation of the feature's importance for a specific business practice (here, churn). The feature importance list signaled the flux of business practices and priorities. If an "intuitive" feature had been sufficiently incorporated in business practices, Daniel and Martin argued, it would – from the model's perspective – cease to be "important." Counter-intuitive findings were *also* valuable. This tension between intuitive results and counter-intuitive discovery showed up again in our interview with Jasper[15]— DeltaTribe's CTO. As he explained, one way to assess the efficacy of a counter-intuitive result was to juxtapose it with the intuitiveness of the data science workflow.

---

[14] In machine learning, *features* refer to measurable attributes. E.g., 'whether email is enabled' is one feature of a customer. Algorithms analyze features to calculate results. A feature importance list, produced by certain algorithms, contains features and their weights. The weights signal the relative importance of each feature in the algorithmic model's analysis.

> "When I get an end-result type of an answer, I like to understand [...] the factors that went into it, and how they [data scientists] weighed it to make sure that it makes intuitive sense. [...] I understand that a lot of the times, findings will be counter-intuitive. I am not immediately pre-disposed to distrust anything that doesn't make intuitive sense. I just like to understand the [significant] factors. [...] Understanding the exact algorithm, probably *less* so. [...] Understanding process is really important [...] to trust that *(pause)* science was allowed to take its proper course" (Interview, November 14, 2017).

The tension between intuition and counter-intuition was not limited to algorithmic results. The business goals themselves, Jasper further argued, may often appear counter-intuitive. This was the case, for instance, with Jasper's requirement of sensitivity over precision. A model configured for sensitivity prioritizes recall, aiming to maximize the number of correct predictions in its results.[16] A model configured for precision, however, prioritizes positive predictive value, aiming to maximize the correctness of its predictions.[17]

> **Jasper (CTO, DeltaTribe):** "As a pragmatist, what I am looking [for] are things that are highly sensitive, and their sensitivity is more important to me than their accuracy. [...] If you can ensure me that of [say] the 2700 [customers] we touch every month, all 500 of those potential churns are in that, that's gold for me. [...] If you could tell me [to] only worry about touching 1000 customers, and all 500 are in it, that'd be even better. But [...], let's start with [making] sure that all the people I need to touch are in that population, and make maximum value out of that. [...] It is about what outcomes I am trying to optimize to begin with, and then what outcomes am I trying to solve for and optimize after. [...] You want your model to be completely sensitive and completely accurate. Of course! [...But,] you don't want to optimize too soon. [...] I probably want to talk about dialing up accuracy a little bit [later]. [Our current approach] is so inherently inefficient that there is an enormous order of magnitude [of] optimization possible without being perfect. [...] It is maybe a little bit counter-intuitive, but the goal I am trying to solve for is: can I spend a subset of my total resources on a population that is going to return well for me, but what I am not doing is avoiding spending resources on people that I should have?" (ibid.)

Business requirements were thus two-fold. On the one hand, the goal was to minimize churn. This led to an initial preference for sensitivity. On the other hand, the aim was to optimize resource allocation. This led to a subsequent preference for precision. For Jasper, the goals came one after the other—first build a sensitive model, and later tune it for precision. The data science team, however, tackled the problem differently. Their models were not configured for sensitivity or precision but for specificity. A model configured for specificity does not focus on maximizing the number or accuracy of its correct predictions but on minimizing the number of its incorrect predictions. The aim was to ensure that healthy customers were not incorrectly identified as likely-to-churn. By minimizing incorrect predictions, the team hoped to ensure that customers classified as likely-to-churn were, in fact, problematic, guaranteeing that resources were not wasted on healthy customers. For Jasper, the two goals married different corporate values with different computational ideals at different points in time. The need to separate the two goals ("you don't want to optimize too soon"), Jasper argued, may seem counter-intuitive to data scientists. Indeed, for the data science team the two goals went hand in hand—the 'specific' solution found in incorrect answers instead of correct ones. At the end of the meeting, business analysts Parth and Charles agreed to conduct pilot tests with the results to see if they lowered the churn rate.

---

[15] Jasper has been DeltaTribe's CTO for 2+ years. Before this, he had 25+ years of work experience in the technology industry in several roles such as chief information officer, chief operating officer, and senior systems analyst.
[16] Recall is the ratio between True Positives and Total Positives. True Positives are customers that churn and are correctly identified by the model as likely-to-churn. Total Positives are the total number of customers that churn. A sensitive model maximizes recall.
[17] Positive predictive value is the ratio between True Positives and the sum of True Positives and False Positives. False Positives are customers that don't churn but are identified by the model as likely-to-churn. A precision model maximizes the accuracy of its predictions.

There are two striking features evident in the vignette above. *First,* quantified metrics such as precision, recall, specificity, and accuracy remain integral to model assessment, but they are often valued differently by different actors. The data science team considered likely-to-churn probabilities a valuable resource, but the business team found them incomplete in the absence of knowledge about how they were generated. Conversely, while the data science team devalued the usefulness of the accuracy score, the business analysts nevertheless considered the score an important signal in assessing model usefulness. The business team wanted to first focus on recall and then on precision, but the data science team saw in specificity a way to marry both concerns. Thus, while different organizational actors differently articulate and negotiate the efficacy of numbers, their use in complex collaborative settings engenders specific forms of practical action (e.g., focusing on varying levels of churn risk as opposed to the binary of churn or no-churn) and understanding (e.g., prioritizing specificity over precision or recall).

*Second,* while intuition acts as a form of a reality check in data science practice, it also runs the risk of naturalizing or rendering invisible biases and precluding alternate findings. Intuitive results and familiar or expected patterns engender trust in algorithmic results, in turn ascribing forms of obviousness to data science approaches. While the role of intuition in making sense of data science results and processes is revealing, this project surfaced another aspect of the role of intuition in data science practice. Organizational actors negotiate not only what is and isn't intuitive, but also when counter-intuition is and isn't warranted. They leverage both the intuitiveness *and* counter-intuitiveness of results and processes to negotiate trust and credibility in data science systems. The possibility of unexpected patterns and counter-intuitive knowledge – one of the great promises of data science work after all – stands in contrast and partial tension with intuitive assessment, requiring ongoing forms of work to balance between the two. This is especially apparent in projects in which business goals themselves may seem to follow a counter-intuitive trajectory to computational ideals.

## 3.2 CASE 2 | SPECIAL FINANCING

For our next case, we turn to a project on *special financing* i.e. loan financing for people who have either low or bad credit scores (usually 300-600) or limited credit history. AutoServe, a DeepNetwork subsidiary, helps people to get special financing loans to buy new and used cars. AutoServe's clientele comprises money lenders and auto dealers who pay AutoServe to receive "leads," i.e. information on people interested in special financing. This information comprises several details including demographics, address, mortgage, and current salary and employment status. Knowing which leads will get approved for loans by lenders/dealers is not straightforward. AutoServe wanted to use data science to predict which leads are likely to get financed *before* they are sent to lenders/dealers. AutoServe assigned Ray[18], AutoServe's business analyst, to work with the data science team on this project.

## (In)credible Data

Both data scientists and business personnel accepted that some leads were good and some bad. For Ray, a lead was good if a lender/dealer approved it for financing. Different lenders and dealers, however, had different requirements for loan approval. The data science team thus approached the question as a matching problem: *how do you match leads to lenders and dealers that are more likely to finance them?* As data scientist Max began work on the project, something troubled him. He described his concern in a data science meeting:

> **Max (data scientist):** Surprisingly, more than 50,000 people who applied for a [special financing] loan earn more than $5000 a month! Why do you *(throws his hands in the air)* need a [special financing] loan for a car if you earn that much money? Maybe there is noise in the data. I need to fix that first (Fieldwork Notes, May 31, 2017).

Max found it odd that people with $5000 monthly earnings applied for special financing. Underlying Max's inference was his assumption that people with bad credit scores do not earn such salaries. Director of data science Martin and DeepNetwork's CTO Justin[19], however, believed otherwise. Arguing that the relationship between credit score and salary was tenuous at best, they told Max that his interpretation was incorrect.

> **Martin (director of data science):** That isn't true. Your current salary cannot tell you whether someone would or wouldn't need a [special financing] loan. Don't fix anything, please!
>
> **Justin (CTO, DeepNetwork):** Yeah, you might have accrued debt or filed for bankruptcy (ibid.).

While Max's concern was dismissed as a false alarm, it did not mean that data was not suspect in other ways. Throughout the project, there were several discussions around credibility ranging from questions of data accuracy to the reliability of the underlying data sources themselves. Business analyst Ray, for example, raised concerns about how data used in the project was generated and collected:

> **Ray (business analyst):** "How this business works […] is very strange. […] The fact that [some] dataset [come] from affiliate[s]. [20] We [also] generate our leads organically [online…], then there are leads that we get from syndication that we bid on. [...] Different pieces of information [are] appended to those leads depending on where […they come] from. […] There is one affiliate […], they give a credit score range for a bunch of those leads. So, not exactly the score, but they could say 'this person is between 450-475 and 525-550,' different buckets. [...] Never realistic, but we pay money and we have this data." (Interview, November 8, 2017).

Acquired through multiple means (e.g., web forms, bidding), lead data was sometimes augmented with additional data, such as credit score ranges, by AutoServe's business affiliates. The role of credit score data was particularly revealing. Credit score ranges were a part of leads bought from a small number of third-party lending agencies. AutoServe's business analysts wondered whether credit score range data could help them in this project. Credit score ranges, for instance, were already used by business analysts as one way to distinguish between two leads that appeared identical but exhibited different financeability.

---

[18] Ray has an undergraduate degree in economics and a graduate degree in applied statistics. He has been with DeepNetwork for 2+ years. Before this, he worked as a Research Assistant (Statistics) in an academic research center.
[19] Justin has been with DeepNetwork for 12+ years (10+ years as CTO and 2+ years as Vice President, Technology). Before this, he had 7+ years of work experience in technology industry in roles such as engineering manager and senior engineer.
[20] The affiliates consist of third-party lending agencies and business associates.

> **Ray (business data analyst):** "Two individuals that have the same age, same income, same housing payment, same everything […] could have wildly different credit scores. [...] You have those two individuals and you send them to the same dealer. From our perspective lead A and B are […] maybe not exactly same but close. [...] But, the dealer can finance person A, and they cannot finance person B [...] So, when they [dealers] are evaluating from month to month whether they would renew their product with us, if we had sent them a bunch from bucket B, and none from A, they are likely to churn. But, we have no way of knowing who is in bucket A and who is in bucket B. […] Two individuals who measure the same on data points, [could] have two different credit scores." (ibid.)

The data science team's attempt to assess the usefulness of credit score range data, however, faced a series of practical challenges. First, the data was incomplete—only available for a few leads (~100,000 out of ~1,000,000). Second, the data was approximate—not in the form of a single number (e.g., 490), but as a range (e.g., 476-525). Third, the data was inconsistent because different affiliates marked ranges differently. For example, a credit score of 490 might be put in the 476-525 range by one affiliate and in the 451-500 range by another. Credit score data, considered an important factor by the business team, was thus at best sparse, rough, and uneven. As data scientist Max attempted to make credit score ranges received from different affiliates consistent with each other, business analysts found a way to make this work easier. Pre-existing market analysis (and, to some extent, word-of-mouth business wisdom) showed that leads with credit scores greater than 500 were very likely to get special financing approval. This piece of information simplified the work of achieving consistency across different credit score ranges. Max did not need to figure out ways to reconcile overlapping ranges such as 376-425 and 401-450 but could simply include them in the *below-500* category. Only two credit score ranges were now important: *below-500* and *above-500*. The solution to the matching problem (*which leads are likely to get financed by a lender*) was now a classification task (*which leads have a credit score of over 500*).

> **Max (data scientist):** [AutoServe] really care about 500. If the credit score is below 500, the dealer will simply kill the deal. They now want me to tell them whether credit score is below or over 500. The problem is there are too many records in 476-525—a very tight group. This makes it difficult (Fieldwork Notes, June 13, 2017).

The presence of several leads in the 476-525 credit score range was a problem given that 500 – a number now central to the project – fell right in the center of the range. This made it hard to figure out which leads in the 476-525 range were above or below 500. The threshold of 500 had helped attenuate the effect of inconsistency but not circumvent it. Max tried several ways to deal with this problem, but each attempt decreased the accuracy of his model. Later, he achieved a classification accuracy of 76%.[21] He did this by completely deleting all the leads from the dataset that were in the 476-525 range. He acknowledged that this was an imperfect solution but argued that it was a good way forward. The model was now accurate for all but a quarter of the leads (better than chance, or 50/50), but at the cost of removing leads near the crucial threshold of 500.

## (In)scrutable Models

When Max shared results with the business team, he received mixed reactions. The business team was unimpressed with the accuracy scores and uncertain about how the model produced results:

> **Bart**[22] **(project manager):** 'We were given a PowerPoint with like, a short note saying here are the numbers and here is this technique. [...] We weren't told much other than that. I personally felt that we weren't evaluating the model, but it was like—do you like these numbers? That wasn't helpful. We didn't like the numbers. [But,] we got no explanation of how things worked, how insights were generated. We all just weren't on the same page. [...] Max tried to explain to us, but it was explained in a manner [...] we maybe didn't get? [...] It soon got overwhelming with all the formulas and models' (Interview, November 1, 2017).

---

[21] The number of correctly identified above-500 leads in the total number of leads that the model identified as above-500.
[22] Bart has been with AutoServe for 4+ years. Before this, he had 14+ years of work experience in the technology industry in several roles such as business development director, technology and design manager, and data analyst.

Max provided detailed descriptions of the model's algorithm, but the "manner" in which he explained did not resonate with the business team. Max had used a neural network. The exact working of a neural network on a specific dataset is particularly black-boxed even for data scientists. In an interview, Max mentioned that the large scale of the data coupled with the complex nature of neural networks made it difficult for him to explain to the business team how the model made decisions. Max, in fact, argued that understanding how the model made decisions was "not that important."

> **Max (data scientist):** The reason we use machine learning, we let the machine learn. Like a child learns a language. When we say 'hi,' the child says 'hi' back. We see that, but we don't know why. We don't ask why the child says 'hi.' I don't get it. We can use tools without understanding the tools. E.g., stock markets. There are charts, lines, and we make decisions based on them. We don't know how they work! (Fieldwork Notes, July 13, 2017).

The data science team was not opposed to explaining the principles underlying their models. It was the *in situ* working of his model that Max argued was black-boxed but also unimportant. In a later interview, director of data science Martin described what he thought was the best way to proceed in such projects—a combination of "implicit trust" and "explicit verification":

> **Martin (director of data science):** "[We need] an implicit trust that the models [produce] correct outputs. […] We can explain at some layman's level […] what algorithms the models are based on, […] what that black box was based on, but please don't ask me – because I cannot tell you anyway – how it got to the result that we are offering to you. I can tell you that […the] model happened to be based on […this algorithm] with 10-fold validation. […] I'll *even tell you* how […the models] are created, what their characteristics are, what [are] the pros and cons of the [model's] algorithm. But, for your case, why did that 0.9 come up versus the 0.84 for customer ID 'x' versus 'z'? Couldn't tell you. […] I am hoping for implicit trust with explicit verification [from a pilot test]. Because if it turns out during the pilot that the effectiveness wasn't there, I am also perfectly okay to call it a failure. It did not meet the business requirements, and we did not add value, and I am okay with that" (Interview, August 23, 2017).

For Martin, implicit trust corroborated by explicit verification in the form of "real-world" tests remedied the lack of explanations. He wanted to conduct a pilot test with model results to see if lender perception of lead efficacy improved. The business team felt otherwise. The absence of model explanations did not inspire enough business trust even for a pilot test. With this feedback, Max returned to work on his model.

Two key insights emerge from the description of this project. *First,* notions of trustworthiness in data science practices are entangled with the perceived credibility (or lack thereof) of data itself. Differentiating between signal and noise in data is challenging. While data scientist Max assumed a set of data values as noise, business analyst Ray doubted the data's reliability given how it was prepared. Incomplete data is a deterrent (e.g., credit score data only available for a handful of leads), but even data assumed complete is often inconsistent (e.g., different ranges provided by different affiliates) or misaligned (e.g., leads in the 475-525 range stood in opposition to treating 500 as a threshold). Working solutions to data-driven problems require creative mechanisms and situated discretion to work *with* messiness (e.g., finding ways to make credit score ranges consistent across affiliates) and *around* messiness (e.g., deleting problematic leads).

*Second*, stakeholders' trust in data science systems stems not only from model results and performance metrics, but also from some explanation or confidence in a model's inner working—an explanation or confidence which may prove challenging to port between members of the data science and business teams. In this project, we see the presence of at least two possible ways to unwrap black-boxed models. One way is to explain *how* a model's algorithm works in principle. Although such explanations are possible and provided by data scientists, we saw in this project (and others) that such information was sometimes considered unhelpful by business teams who preferred a second kind of explanation: *why* the model makes a specific decision. Such explanations are neither straightforward nor always available. Data scientists describe the lack of these explanations not as an impediment but as a trade-off between in-depth understanding and predictive power. It is thus not surprising that data scientist Max considered such explanations unnecessary. The absence of these explanations necessitates additional work on the part of the data science team to help foster business trust in black-boxed models (e.g., combining implicit trust and explicit verification).

## 4 DISCUSSION

The two cases above surface crucial tensions within real-world data science practices. In case one – *churn prediction* – we see negotiations concerning the (un)equivocality of numbers and the (counter)intuitiveness of results. Quantified metrics, integral to assessing the workings of data science solutions, exhibit a certain degree of plasticity. The perceived value of metrics shifts as numbers move between people and teams marked by different, often divergent, valuation practices. Recourse to intuition may engender confidence but at the risk of camouflaging or dismissing novel insights. Balancing between expected and unexpected results is therefore central to the validity of intuitive assessments. Through case two – *special financing* – we see how actors assess data (in)credibility and rationalize model (in)scrutability. Prior knowledge and established goals shape a dataset's perceived effectiveness, requiring discretion to work with and around challenges of inconsistency, unreliability, and incompleteness. Explaining why models make certain decisions (i.e., their situated application) is often as important as describing how they work (i.e., their abstract working). The (in)scrutability of a model shapes its evaluation in significant ways. These tensions problematize the actors' ability to trust data, algorithms, models, and results. In the face of uncertainty, we see actors using specific mechanisms to resolve and circumvent such problems—solutions to problems of trust that help enable pragmatic action. We describe these mechanisms in the following four sub-sections.

**Contextualizing Numbers**

In case one, we see actors qualify the effective value of quantified metrics in specific ways. A first strategy involves *decomposing a problematic number into its constituent parts*. For the data science team, a single accuracy score made invisible, especially to business teams, the four scores constituting it (true positives, true negatives, false positives, false negatives). An overall accuracy score of, say, 75% demonstrates a model's success for three-quarters of the data but signals the model's failure for the remaining quarter. Decomposing the score pluralizes the notion of a mistake, recasting the missing 25% as a combination of four *different kinds* of mistakes embedded within statistical ideals of precision, recall, or specificity. The business team wanted to first build a sensitive model and then tune it for precision. The data science team, however, focused on specificity, trying to kill two birds with one stone—minimizing incorrectness to improve recall as well as sensitivity. Different approaches enable prioritization of certain mistakes and values over others, facilitating the re-negotiation of the model's perceived success or failure. During fieldwork, we saw several instances in which numbers considered sub-optimal were broken down into their constituent parts, while numbers assumed adequate or sufficiently high were often communicated and interpreted at face value.

A second strategy involves *situating suspect numbers in a broader context.* The accuracy score provided information about model performance, but the score's interpretation extended beyond the model. The data science team juxtaposed low accuracy scores with the description of churn prediction as a "very difficult" problem. Arguing that customers often churn for reasons uncapturable in datasets, they found it unreasonable to assume that human behavior could be modeled perfectly. Low performing models shaped the data science team's understanding of the project's complexity. The value of "even bad results or less than ideal results" was found in their ability to provide information not already available to the business team. The accuracy score of 30 was not 70 less than 100 but, in fact, 30 more than zero. Sub-optimal models were still better than nothing. Throughout our fieldwork, we saw instances in which numbers, especially large numbers, acted as "immutable mobiles" [60]—as stable and effective forms of data science evidencing. But in many others, actors leveraged the inherent mutability of numbers in specific and significant ways. The plasticity of numbers in such contexts is therefore partial but strategic.

**Balancing Intuition**

Actors balance intuition and counterintuition in specific ways. A first strategy comprises *leveraging intuition as an informal yet significant means to ratify results and processes*. We saw this mechanism at play in case one when data science team members inquired whether computed feature importances matched existing business insights. The convergence between model results and prior knowledge engendered confidence in the models even when their inner workings were not available for inspection. The model's capability to "discover" already-known business facts inspired trust in its analytic ability. In addition to endorsing results, intuition aids assessing project workflow. In case one, and multiple times during our fieldwork, we saw business actors enquiring into data science processes with an explicit intent to ensure that followed protocols "made intuitive sense"—to the extent that the intuitiveness of data science workflows was considered a way to assess the efficacy of counter-intuitive results. Different from the scrutiny data scientists already employ in their everyday work (e.g., preventing model overfitting), such examinations were considered a way to uncover erroneous decisions (e.g., data reorganization) or configurations (e.g., flawed assumptions).

Deploying intuition as a form of assessment, however, has its own problems. Intuitive results call further attention to the subset of counter-intuitive results. When certain feature importances matched business expectations, business analysts questioned the absence of other expected features. Upholding the validity of intuitive assessment in such situations required a way to explain counter-intuitive findings while not entirely relinquishing the ability for intuitive ratification. This was achieved through a second strategy: *demarcating between algorithmic and human analytical approaches to justify perceived differences.* Data science team argued that, unlike humans, algorithms statistically traverse the uneven contours of data, producing results that may sometimes appear unrecognizable or different. Counter-intuitive findings, they argued, can at times comprise novel forms of knowledge and not model mistakes. Balancing between intuition, counter-intuition, and trust requires work on the part of organizational actors to negotiate the relations between prior knowledge and novel discoveries. Often, intuitive results are made visible to inspire trust in models, while sometimes counter-intuitive results are posited as moments of new discoveries. The excessive overlap between model results and prior expectations is also problematic at times since intuitive results can stem from overfitted or over-configured models. For example, in a different project, we saw that a model whose results *completely* mirrored existing insights was deemed a failure by business personnel who argued that the model "lacked business intelligence" because it furnished "no new information."

**Rationalizing and Reorganizing Data**

In case two, we see how actors negotiate and accomplish data credibility in at least two different ways. A first mechanism involves *rationalizing suspect data features*. The data scientist questioned the high salary figures for certain customers with low/bad credit scores. (Differentiating between high/low salaries is itself a matter of interpretation). Assuming that people with low/bad credit do not earn high salaries, he wanted to get rid of this data. Business personnel, however, invoked the fragile relationship between fiscal circumstances and monthly earnings, arguing that people with seemingly high salaries were not atypical but ordinary occurrences in the world of special financing. Such a form of rationalization involved the contextualization of data in prior knowledge to articulate felt, yet practically oriented, experiences of data inconsistency, unreliability, and illegitimacy. Technical examination and statistical measures are highly visible forms of data credibility arbitration within data science. In this case, and many others, however, we saw that the lived experience of data is a significant yet largely under-studied form of credibility assessment.

A second mechanism comprises *reorganizing data in different ways to mitigate identified problems*. Throughout the project, several problems with the special financing dataset were identified such as issues of consistency (credit ranges varied across affiliates) and interpolation (leads in the 476-525 range needed approximation around the 500 threshold). Nevertheless, such issues did not obstruct the project. Identified problems were tackled in specific, often creative, ways. The data scientist tried several ways to achieve compatibility between inconsistent credit score ranges. The effect of inconsistency was lessened by the fact that business analysts characterized the credit score of 500 as a significant cutoff (scores greater than 500 were considered highly likely to get special financing approval). There was no need to make all divergent ranges compatible. The leads could simply be restructured into two buckets: above-500 and below-500. Leads in the 476-525 range, the border between the two classes, were now a significant problem. Placing these leads in either bucket required work and discretion to interpolate scores in some manner. The problem was resolved by expunging all the leads in the 476-525 range—a solution considered imperfect but practical.

**Managing Interpretability**

In case two, two kinds of explanations were discussed: how a model works (i.e., the abstract working of the model's underlying algorithm) and why a model makes specific decisions (i.e., the situated application of the model on a data point). Explaining the in situ working of, for instance, neural networks is difficult and often impossible. The decision process of even relatively simpler models is hard to grasp for large-scale data. Data scientists focused instead on the abstract working of the model's underlying algorithm. Few business personnel (particularly those with technical backgrounds) found such explanations useful. The majority considered them impractical, wanting to understand the rationale behind model decisions more specifically or model complexity more generally. Business personnel's ability to trust results, as they repeatedly told us, was severely affected when faced with such forms of "opaque intelligence" [97].

At several instances during our fieldwork, and as described in case two, data science team members tried to alleviate this problem by *accentuating the perceived import of model results,* in turn *deemphasizing the need to understand algorithms and models.* Business personnel desired predictive prowess *and* analytic clarity. Data scientists argued for a trade-off between understandability and effectiveness—state-of-the-art models were not entirely inspectable. As one data scientist said, the complexity of models is not a problem but the very reason why they work—a resource for model performance instead of a topic for analytic concern. Transparency remained a problematic ideal caught between multiple interpretations of inscrutability [64]. Opacity was often perceived as a function of models' black-boxed nature, necessitating detailed descriptions of algorithmic workings. Even when translucent, models remained recondite—their workings were complex; their results were hard to explain. Underscoring the import and value of results in these circumstances, deemphasized complex descriptions and absent explanations. The question changed from how or why models worked to whether or how well they worked. "Implicit trust" took the place of complex descriptions. "Explicit verification" from real-world tests supplanted absent explanations. What remained unresolved, however, was the foreign nature of algorithmic approaches themselves. Models were opaque and abstruse, but also alien—their complexity was described and explained, but not justified [86].

**Collaboration, Translation, and Accountability: Implications for CSCW Research and Practice**
These findings hold important implications for the growing data science field, both within and beyond CSCW. Trustworthy data science systems are a priority for organizations and researchers alike—evident, for instance, in rubrics for assessing a data science system's production-readiness [13] or rules for conducting responsible big data research [105]. Such forms of advice address a range of sociotechnical challenges, helping data scientists manage aspects ranging from performance evaluation and feature engineering to algorithmic harm and ethical data sharing. As CSCW and critical data studies researchers work to make data science approaches transparent, metrics humanistic, and methodologies diverse, their tools often travel far from their academic and research contexts of development, finding new homes in company servers, business meetings, and organizational work. The insights in this paper add three further dimensions to the effective practice and management of data science work by explicating how specific tensions problematize trust and credibility in applied data science systems, and how these problems are variously negotiated and resolved.

*First,* rather than a natural or inevitable property of data or algorithms themselves, the perceived trustworthiness of applied data science systems, as we show in this paper, is a collaborative accomplishment, emerging from the situated resolution of specific tensions through pragmatic and ongoing forms of work. As our actors repeatedly told us, data are messy and models only approximations (or as the classic line on models has it: "all models are wrong, but some are useful" [11]). Perfect solutions were not required—models simply needed to be "good enough" for the purpose at hand [49]. Actors' trust in data science did not therefore depend on the flawless nature of its execution rather on the reasoned practicality of its results. Much like actors in a "heterarchy" [94], we saw organizational actors treat everyday uncertainties less as impediments and more as sites for justifying the "worth" [8] of data, models, and results through actionable strategies. Organizational actors often acknowledged the always-already partial and social nature of data and numbers. Their attempts to negotiate and justify the worth of data science systems were thus aimed at identifying pragmatic ways to make the "best" out of inherently messy assemblages. Such uncertain moments comprise forms of focused skepticism—doubt in and negotiation of specific aspects of data science work (e.g., counter-intuitiveness) require trust in several other aspects (e.g., data sufficiency, model efficacy). This further speaks to the intimate relationship between trust, skepticism, and action: or, as Shapin [88:19] argues, "distrust is something which takes place on the *margins* of trusting systems."

Data science, particularly from an academic or research perspective, is often imagined from the outside as the work of data scientists interacting around reliable and widely shared tools, norms, and conventions—a "clean room" imagination of data and its relationship to the world. As our cases show, however, corporate data science practices are inherently heterogeneous, comprised by the collaboration of diverse actors and aspirations. Project managers, product designers, and business analysts are as much a part of applied real-world corporate data science as are data scientists—the operations and relations of trust and credibility between data science and business teams are not *outside* the purview of data science work, but *integral* to its very technical operation. A more inclusive approach to the real-world practice of corporate data science helps us understand that while quantified metrics and statistical reasoning remain visible and effective forms of *calculated trust*, the crystallization of trust in applied data science work is both calculative *and* collaborative. Quantified metrics allow close inspection of data and models, yet numbers appear differently to different actors—sometimes stable, and at other times mutable. Numbers not only signify model performance or validity, but also embody specific technical ideals and business values. Understanding the pragmatic ways of working *with* the plastic and plural nature of quantified trust and credibility metrics can further nuance existing CSCW and HCI research on the design of trustworthy systems [37,58,82] and reliable performance metrics [1,48,79]—managing numbers is as important as engineering them.

*Second,* we show how the collaborative accomplishment of trust requires work on the part of diverse experts to translate between different forms of knowledge. For instance, data scientists work to explicate algorithmic approaches to business analysts, and business teams strive to explain business knowledge to data scientists. We see in such forms of translation work a common trait—the recourse to stories and narratives to not only explain algorithmic results [30,98], but also describe suspect data and model attributes. Narrativization serves various purposes ranging from delineating the abnormal from the ordinary (e.g., what is and isn't noisy data) to rendering opaque technicalities natural and commonplace (e.g., models are inscrutable, but so are human brains). As exercises in world-making [41], narratives invoke specific lifeworlds [39,44] to explain what things mean or why they are a certain way. Algorithmically-produced numbers may be black-boxed, but tales about and around such numbers engender forms of intuitive and assumptive plausibility. While datasets comprise information on people and their practices, people remain largely invisible in data, dismembered into rows, columns, and matrices of numbers. Forms of narration and story-telling, however, are often all about people, significantly shaping their identity, agency, and forms of life. Narrativization, as a form of doing, implicates data science between reality and possibility, between signal and noise—indeed, between life and data.

These insights on the narrativization of data and results add new dimensions to existing CSCW and HCI research on explainable machine learning systems [1,82,86,101] and human perception of data representations and algorithmic working [21,48,59,103], making visible not only the plurality of reasonings and modes of justification [8] that actually subtend applied data science work, but also the multiple forms of expertise that constitute such work in complex real-world settings. Workable data, for instance, is computationally accomplished through multiple forms of pre-processing—each attempt at reorganization adds value, but also removes possibilities. Researchers strive to create better tools to identify and resolve issues with real-world data, but even data assumed or made workable by data scientists are sometimes distrusted by other organizational actors. The identification of mess is an exercise not just in statistics and computation, but also narrativization and interpretation. Understanding and articulating the relation between distinct forms of data curation and their interpretational and narrative affordances, for instance, can complement current technical work on data pre-processing and curation—artifacts that are simultaneously partial and practical. Imagined as an exclusively algorithmic venture, data science would appear as the stronghold of data scientists working with specialized and sophisticated computational tools. Acknowledging the domain-specificity of data, however, surfaces the many other forms of necessary expertise supplied by diverse organizational actors. These different experts influence the development of a data science system in different ways, pulling the project in specific, sometimes contradictory, directions. Understanding the work of these experts can provide new pathways for CSCW, HCI, and critical data studies researchers into the study, design, and management of data science systems.

*Third,* we show how different experts hold accountable different parts of data science systems, highlighting the distributed and multifarious nature of data science trustworthiness. For instance, data scientists cross-validate results, while business analysts inquire about data scientists' business knowledge and assumptions. In some cases, trust is placed not in the analysis, but in the identity of the analyst. At a few points in our fieldwork, we saw that data and results were assumed correct by business stakeholders because of the trust they placed in specific individuals. Data scientists trusted datasets more if they came from business analysts as opposed to data engineers. Business teams trusted results coming from senior analysts and scientists. As forms of 'social scaffolding,' people's perceived reputation and knowledgeability at times provided working solutions to problems of credibility. Embedded within these different forms of trust valuations, are distinct approaches to data science auditing. On the one hand, audits function as a form of *algorithmic witnessing*— backtracking technical procedures to ensure the reliability and validity of model results. As exercises in corroboration, such audits necessitate data science expertise on the part of the auditors. On the other hand, audits contribute to *deliberative accountability* [46]—situating model contingencies, technical discretion, and algorithmic results in the broader social, cultural, and organizational milieu. Acknowledging the role of other experts, such audits encompass multiple ways of 'seeing'—of witnessing data science results and processes. Between *algorithmic witnessing* and *deliberative accountability*, we see the everyday work of applied corporate data science betwixt and between algorithms, businesses, calculations, and aspirations—technically valid, but also practically sound.

Juxtaposing algorithmic witnessing with deliberative accountability provides new research pathways into the effective evaluation, governance, and management of data science systems. As CSCW and HCI researchers work to make data science models transparent and results explainable, their focus should include not only unpacking algorithmic black-boxes, but also studying how data and models move between teams, products, and services. This enables us to better understand how the inability of organizational actors, sometimes even of data scientists, to understand models is an artifact not only of their black-boxed nature, but also, for instance, of their counter-intuitiveness. This is especially problematic given that a large part of data science's appeal is its ability to surprise us. Even when opened and made tractable, model innards and results remain complex and foreign in their movement between people, practices, and teams. Like other forms of alternate knowledge, data science's alien-ness complicates the attribution of trust and credibility to unaccountable and inscrutable truths—its foreignness sometimes mistaken by some for its incorrectness, and at other times celebrated as a novelty. As researchers make visible the rules comprising models and describe the application of these rules, they also need ways to explain and unpack the complex and alien nature of the rules themselves: "while there will always be a role for intuition, we will not always be able to use intuition to bypass the question of why the rules are the rules" [86].

Our findings also suggest some more immediately practical takeaways for data science work in the contexts of academic education, organizational practices, and data science research more generally. In learning environments, would-be data scientists are encultured into data science's professional discourse, largely comprising of technical work such as data pre-processing, model selection, and feature engineering. As students go on to take data science jobs in corporations, their everyday work, however, comprises working not only with data and models, but also with project managers and business stakeholders. The collaborative and heterogeneous nature of real-world data science work remains as of now largely invisible in current data science curricula and training. Several of our actors argued that the data scientists they interacted with lacked, among other things, the "vocabulary" of working with business teams. As James – a senior business team member – put it: "data scientists […are] very eager to crunch numbers […], train the system, and see what kind of output they […can] get" (Interview, 6 November 2017). The incorporation of collaboration (e.g., interacting with non-data-scientists) and translation (e.g., effective communication of results) work into data science curricula and training is thus a good first step to ensure that would-be data scientists not only learn the skills to negotiate the trust in and credibility of their technical work, but also learn to see such forms of work as integral to the everyday work of data science. Or, to put it in terms of sociologists Harry Collins and Robert Evans, real-world applied data science projects require forms of both "contributory" *and* "interactional" expertise [18].

In corporate organizations, as we show in this paper, the development of data science systems comprises a combination of algorithmic and deliberative accountability. Integral to both approaches is the need and role of documentation. At the organization, we initially discovered that there was much emphasis put on code documentation, but the everyday discretionary work of data scientists in pre-processing data, selecting models, and engineering features remained less visible and documented (we attempted to address this at the organization by initiating detailed project and decision documentation). This remains a hindrance not only for forms of inter-organization accountability, but also for the compliance and management of such systems in the wake of laws and policies such as the General Data Protection Regulation (GDPR) and the Right to Explanation in Europe. With current calls for more open documentation, corporate organizations need to document not only algorithmic functions and data variables, but also data decisions, model choices, and interim results. Organizations need to allocate additional resources and efforts to make visible and archive the seemingly mundane, yet extremely significant, decisions and imperatives in everyday data science work.

Lastly, highlighting the existence and role of diverse experts in applied data science projects, our work helps to further unpack the distinction between the designers and users of data science systems in existing CSCW and critical data studies research. The study of in-use public-facing data science products often works on a clear distinction between designers and users (e.g., Google data scientists made Google Search, which is now used by internet users). Unpacking the design and development work of such corporate systems, however, stands in contrast with rigid binaries— corporate organizations are not monolithically comprised of data scientists working with their algorithmic toolkits to produce computational artifacts. Project managers, product managers, and business stakeholders, as we show, aren't merely the "users" of data science systems, but also in-part their managers, stakeholders, and designers. Interpretability remains multiple [64], but so do the people requiring explanations. As we focus on studying and building interpretable models and trustworthy systems, we must also consider *who* attributes trust and requires explanations, *how,* and for *what* purposes. The decision to deploy a data science solution in a real-world product remains a negotiated outcome—one in which data scientists play an important, yet partial, role.

**Limitations and Future Research**

The specific nature of the dynamics witnessed in our case may differ across organizations, depending on the varying ways and standings with which data science (and data scientists) are incorporated into broader corporate settings and goals. Is data science perceived an asset to organizational knowledge because of its analytic capabilities or deemed necessary because of market competition? Do data science team members have a lot to lose if things do not work or are they allowed to experiment and make mistakes? Who has the final word on data science projects—data scientists, project managers, business analysts, or business executives? Do data scientists work across business verticals or are they assigned to specific areas of business practice? How do the differences in the educational and professional backgrounds of data scientists, project managers, and business analysts impact data science work? Such questions speak to the uncertainties and heterogeneities attending the entry and practices of applied data science in corporate environments, and the need for further empirical study.

## 5 CONCLUSION

In this paper, we showed how tensions of (un)equivocal numbers, (counter)intuitive knowledge, (in)credible data, and (in)scrutable models shape and challenge data science work. We described a range of strategies (e.g., rationalization and decomposition) that real-world actors employ to manage, resolve, and sometimes even leverage the problems of trust emanating from these tensions, in the service of imperfect but ultimately pragmatic and workable forms of analysis. As we work to guarantee and foster trust in applied data science, understanding the situated and discretionary work of trust and credibility negotiations generates new possibilities for research and practice—from the implications of narrative affordances and plurality of model opacity to the management of numbers and leveraging of expertise. As data science grows, so do its problems of trust. This paper is our attempt to highlight that trustworthiness in organizational data science work emerges through a mix of algorithmic witnessing and deliberative accountability—a calculated, but also a collaborative accomplishment. The organizational work of data science comprises not just pre-processing and quantification, but also negotiation and translation. It is these processes together, and not any taken singly, that ultimately accounts for the trustworthiness and effectiveness of data science.

## ACKNOWLEDGMENTS

The funding for this research was provided by National Science Foundation grant CHS-1526155. We wish to thank our anonymous reviewers for their feedback, and Solon Barocas, David Mimno, members of Cornell University's Culturally Embedded Computing (CEMCOM) research group, and participants of the Lives of Data 2.0 SARAI workshop (January 5-6, 2018) for comments and suggestions on earlier versions of this work.

26  S. Passi and S. J. Jackson